\documentclass[journal,letterpaper]{IEEEtran}

\usepackage{amsmath}
\usepackage{amssymb}
\usepackage{amsfonts}
\usepackage{algorithm}
\usepackage{graphicx}
\usepackage{epsfig}
\usepackage{subfigure}
\usepackage{psfrag}
\usepackage{cite}
\usepackage{color}

\usepackage[letterpaper,margin=0.8in]{geometry}
\usepackage{dblfloatfix}

\begin{document}
\title{A D2D-based Protocol for Ultra-Reliable Wireless Communications for Industrial Automation} %: \\ Cost of User Activity Detection on User Rates}
\author{Liang Liu, \IEEEmembership{Member,~IEEE} and Wei Yu, \IEEEmembership{Fellow,~IEEE}
%, \IEEEmembership{Member,~IEEE} and Wei Yu, \IEEEmembership{Fellow,~IEEE}

\thanks{Manuscript received October 14, 2017, revised January 28, 2018 and March 6, 2018, accepted April 26, 2018. The associate editor coordinating the review of this paper and approving it for
publication was Dr. Lin Dai. This work is supported by Nokia Bell Labs, NJ, USA, and by Natural Sciences and Engineering
Research Council (NSERC) of Canada.}
\thanks{The authors are with The Edward S. Rogers
Sr. \mbox{Department} of Electrical and Computer Engineering, University of Toronto,
Toronto, Ontario, Canada, M5S3G4,
(e-mails:lianguot.liu@utoronto.ca; weiyu@comm.utoronto.ca).}}
%\thanks{This work is supported by the Natural Sciences and Engineering Research Council (NSERC) of Canada through a Collaborative Research and Development (CRD) grant.}}

\maketitle

\begin{abstract}
As an indispensable use case for the 5G wireless systems on the roadmap, ultra-reliable and low latency communications (URLLC) is
a crucial requirement for the coming era of wireless industrial automation. The key performance indicators for
URLLC stand in sharp contrast to the requirements of enhanced mobile broadband (eMBB): low-latency and ultra-reliability are paramount
but high data rates are often not required. This paper aims to develop communication techniques for making a paradigm shift from the conventional
human-type broadband communications to the emerging machine-type URLLC. One fundamental task for URLLC is to deliver
short commands from a controller to a group of actuators within the stringent delay requirement and  with high-reliability. Motivated by the
factory automation setting in which tasks are assigned to
groups of devices that work in close proximity to each other thus can form clusters of reliable device-to-device (D2D)
networks, this paper proposes a novel two-phase transmission protocol for achieving URLLC. In the first phase, within the latency requirement, the multi-antenna base station (BS) combines the messages of all devices within each group together
and multicasts them to the corresponding groups; messages for different groups are spatially multiplexed. In the second phase, the devices that have decoded the messages
successfully, herein defined as the leaders, help relay the messages to the other
devices in their groups. Under this protocol, we design an innovative leader selection based beamforming strategy at the BS by utilizing sparse optimization technique. The proposed strategy leads to a desired sparsity pattern in user activity with at least one leader being able to decode its message in each group in the first phase, thus ensuring
full utilization of the reliability enhancing D2D transmissions in the second phase. Simulation results are provided to show that the proposed two-phase transmission protocol
considerably improves the reliability of the entire system within the stringent latency requirement as compared to existing schemes for
URLLC.
\end{abstract}

\begin{IEEEkeywords}
Ultra-reliable and low latency communications (URLLC), 5G, industrial automation, device-to-device (D2D) communications, machine-type communications (MTC), multicasting, beamforming, sparse optimization.
\end{IEEEkeywords}

\IEEEpeerreviewmaketitle

\newtheorem{definition}{Definition}
\newtheorem{example}{Example}
\newtheorem{theorem}{Theorem}
\newtheorem{proposition}{Proposition}
\newtheorem{remark}{Remark}
\newcommand{\mv}[1]{\mbox{\boldmath{$ #1 $}}}

\section{Introduction}\label{sec:Introduction}
Enhanced mobile broadband (eMBB), massive machine type communications (mMTC), and ultra-reliable
and low latency communications (URLLC) are the three main use cases that the 5G technology must support \cite{3GPP}.
Addressing the above requirements in 5G calls for new methods and ideas at both the component
and architectural levels, including massive multiple-input multiple-output (MIMO) \cite{marzetta,larsson14},
millimeter wave (mmWave) communications \cite{Roh14}, and cloud radio access network (C-RAN) \cite{Simeone16}
for eMBB, as well as multiple access schemes to support a massive number of devices for mMTC
\cite{Liu17_massive1,Liu17_massive2,Bjornson}. This paper focuses on URLLC. Specifically, we aim to tackle the latency and reliability
requirement motivated by industrial automation applications \cite{Kristensen14,Frotzscher14,Varghese15,Jiang16,Chen16}.

In a typical closed loop industrial control scenario, groups of sensors and actuators are deployed in a fixed area in a factory setting.
Periodically or when triggered by external events, the sensors send their measurements to the central controller, which then makes decisions and
sends commands to the actuators for action. Under the current technology, sensors and actuators are typically connected to the central controller via
a wired configuration in most factories. In the near future, under the fourth industrial revolution roadmap (known as Industry 4.0), the communication networks in the factory setting are expected to migrate from wired to wireless for the purpose of increasing the flexibility in
moving machinery and also for reducing the infrastructure expenditure \cite{Kristensen14,Frotzscher14,Varghese15,Jiang16,Chen16}.
As factory automation systems are highly sensitive to signal
delays or distortions, such a transition will impose challenging requirements in terms of latency
as well as reliability for the wireless technologies. In the current 4G cellular network, the end-to-end latency (which includes data transmission, packet retransmission, signal processing, protocol handling, and switching and network delays) can be in the order of 30-40ms, with the physical-layer latency accounting for about 15-20ms. For mission-critical applications for industrial
automation, the
latency requirement of 5G physical layer is expected to be pushed down to less than $1$ms, an order of magnitude shorter than 4G. Further, such low latency requirement needs to be satisfied with
ultra-reliability, e.g., $99.999\%$ or higher.

%\begin{figure}
%\begin{center}
%\scalebox{0.6}{\includegraphics*{wireless_sensor_and_actuator_network.eps}}
%\end{center}
%\caption{Closed loop industrial control: sensors send measurements to the controller in the uplink; the controller sends commends to actuators in the downlink.}\label{fig1}
%\end{figure}

This paper aims to address the challenge of wireless factory automation by focusing on the downlink URLLC
in one cell (factory) of a cellular system, where the multi-antenna base station (BS) (the central controller)
needs to send a small amount of information bits (command) to each user (actuator) within the latency requirement ($1$ms). The core question this paper tries to answer is how to achieve the above goal
with ultra-reliability in the sense that all the users can decode their messages with a very high probability.

Achieving URLLC with the conventional broadcasting strategy is difficult due to the fact that a typical factory may have hundreds (or even thousands) of actuators.
Specifically, in a massive connectivity scenario where the number of devices is
larger than the number of antennas at the BS, it can be difficult to transmit at an appreciable data rate to each user reliably,
especially for the cell-edge users that suffer from strong inter-cell interference. Moreover, time-division multiple-access (TDMA) may not be a feasible strategy
because each user would only be allocated a fraction of the total transmission time, then the required signal-to-noise ratio (SNR) to achieve the target rate would be very high.

\begin{figure}
\begin{center}
\scalebox{0.48}{\includegraphics*{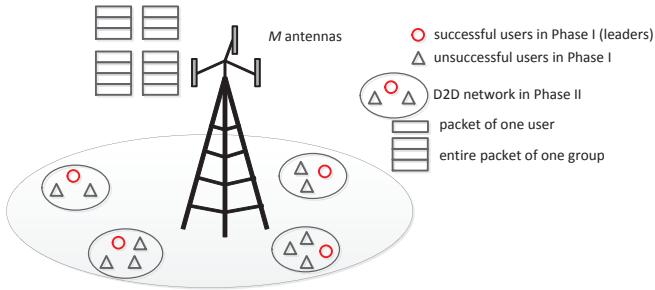}}
\end{center}
\caption{Illustration of the two-phase transmission protocol: in the first phase, BS combines each group's messages together and multicasts them to the leaders in the corresponding groups; in the second phase, each leader helps relay the messages to the unsuccessful users in its group via the D2D network.}\label{fig2}
\end{figure}

This paper proposes a novel transmission strategy for URLLC that relies on a key observation that in practice, devices in a factory setting, e.g., robots or 3D printers, typically work in close proximity to each other and thus can potentially form a
device-to-device (D2D) network for peer-to-peer communications. It is envisioned that the communication
within each D2D network is significantly more reliable than that from the BS to the users due to the much stronger channels between
the users in the same group. To exploit the reliable D2D networks, this paper proposes a novel D2D-based two-phase transmission protocol
as shown in Fig. \ref{fig2}, in which the BS sends the messages to the users in the first phase, while in the second
phase, the users who have already decoded the messages successfully (defined as the \emph{leaders} of the groups) help relay the
information to the other users in the same groups who have failed to receive their messages previously.
Note that the reliability of the overall system is limited by the reliability of the cell-edge users.
Our proposed protocol can opportunistically activate the cell-edge users who happen to not suffer from
strong inter-cell interference due to the channel fading and let these leaders help the other
cell-edge users with low signal-to-interference-plus-noise ratio (SINR) to achieve high reliability in the second phase.

To enable the proposed relay strategy to work, the leaders of each group need to receive the entire messages for all the users
in their group. This paper devises a multi-group multicasting technique in the first phase
\cite{Sidiropoulos08}, in which the user messages in each group are combined together as a single message and multicast to the leaders in the
corresponding group, while messages for different groups are spatially multiplexed. Such a message combination strategy typically results in a manageable multicasting rate,
since in most URLLC scenarios each device only requires a very small number of information bits within the latency requirement
such that the total rate over all users in the same group is still reasonably small.
Moreover, a substantially smaller number of users need to be activated
in Phase I as compared to the broadcasting scenario since one group only needs one leader.

Since the users in the same group usually belong to the same factory, the incentive mechanism and security, which are
challenging issues in practical D2D networks \cite{Kadowaki15}, are no longer the main considerations in our investigated setup. Instead, leader selection in the
first phase becomes the deciding factor of our protocol, since the groups without leaders cannot utilize the reliable D2D networks
in the second phase. This paper proposes a dynamic leader selection based beamforming solution based only on the
instantaneous downlink channel state information (CSI) (without needing the CSI of the D2D networks) such that at the end of the first phase
each group has at least one leader with high probability.

\subsection{Prior Work}
The wireless inter-connection of the traditional manufacturing industries is a crucial goal for future wireless standards \cite{Kristensen14} -- \cite{Chen16}.
The current wireless techniques are not designed for the stringent reliability and latency requirements of the mission-critical applications.
As a result, designing new techniques for URLLC is considered as an increasingly important
goal for 5G \cite{Popovski15,Popovski16}, with some initial efforts already taking place.
For example, \cite{Johansson15} provides a high-level discussion about the potential to utilize diversity, e.g., MIMO, convolutional codes, and hybrid automatic repeat
request (HARQ) scheme \cite{Shariatmadari15} to achieve URLLC. Moreover, coordinated multi-point (CoMP) \cite{Ohmann14}, deployment strategies such as adjusting the cell size \cite{Brahmi15},
adaptive modulation and coding (AMC) \cite{Shariatmadari16}, as well as reduced transmission time intervals and shorter symbol durations in orthogonal frequency-division
multiplexing (OFDM) systems \cite{Yilmaz15} are also investigated to improve the reliability of wireless communications. However, these works in general are built on the
traditional wireless techniques that are mainly driven by the broadband communications and are often inefficient for URLLC. There is general consensus that some fundamental
change in the transmission protocols is necessary to satisfy the stringent latency and reliability requirements imposed by the future wireless industrial automation
\cite{Popovski16}.

A recent work \cite{Sahai15} presents an interesting two-phase transmission protocol, named {\it Optimizing Cooperative Communication for Ultra-reliable Protocols Yoking Control Onto Wireless} (Occupy CoW), for both the uplink and downlink URLLC,
that makes use of cooperative relaying to reach very high levels of reliability, while maintaining a fixed cycle time of $2$ms in a network of $30$ nodes.
For the downlink communication, specifically, the BS combines all the users' messages together and multicasts them to the users in the first phase, while the users that
can decode the messages help relay them to the other users in the second phase. However, if there are too many users in the system, such a combination of
all users' messages may lead to a very high multicasting rate, resulting in too few leaders in the first phase. Our work builds upon the Occupy CoW protocol but differs in
the sense that the geographic information of the users is utilized to divide them into groups: only the messages of each group is sent
to the corresponding leaders, as each leader can subsequently help its neighbors. One obvious advantage of such a grouping strategy is
a lower multicasting rate for each group, instead of a higher multicasting rate across all the users in the network.

It is also worth noting that the conventional approach for sending individual messages to users in the downlink is information broadcasting. In the case of a
single-antenna BS, the joint power and admission control problem for such a setting is investigated in \cite{Mitliagkas11,Luo13}, in which the number of users achieving their SINR
targets is maximized in the event that not all of them can achieve their SINR targets (e.g., when the SINR feasibility condition for power control defined in \cite{Foschini13} does not hold). However, the goal
for URLLC is to provide reliable services to all the devices, rather than a subset of devices. As a result, our work can be interpreted as an effort to enlarge the feasible
SINR regime of the conventional power control and beamforming technique \cite{Foschini13} by a utilization of the D2D network such that URLLC can be achieved in
more challenging settings.

\subsection{Main Contributions}
The main contributions of this paper are summarized as follows. First, this paper proposes a novel two-phase transmission protocol for URLLC based on the observation that a group of devices in close
proximity to each other can form a D2D network in which reliable communication is possible. Under the proposed protocol,
the BS combines each group's messages together and multicasts them to the corresponding groups in the first phase, while the users that decode the messages
successfully, i.e., leaders, help relay the messages to the other users in the same group in the second phase. We point out that since reliable
communication is possible in each D2D network, the core issue under our proposed protocol is the beamforming design in the first phase that aims to successfully transmit to at least one leader in each group, while the other users can rely on the leaders in the second phase.

Second, we formulate the leader selection based beamforming problem in the first phase from a sparse optimization perspective by introducing a set of auxiliary variables
that indicate the gap between each user's SINR and its SINR target. Such a formulation enables us to design
the beamforming at the BS and select the leaders of each group jointly rather than separately. Moreover, leader selection results in a new and non-trivial
sparsity pattern for the auxiliary variables since in each group at least one auxiliary variable should be zero (which implies zero gap between the SINR and SINR target
and thus a leader). To achieve this desired sparsity pattern, we introduce a novel geometric-mean based penalty for the auxiliary variables
of each group, which is minimized to zero when each group has at least one leader. Numerical results are provided to show that such a penalty guarantees a
fair leader assignment among groups.

Finally, we provide a comprehensive performance comparison between our proposed strategy and the existing ones in the literature, e.g., Occupy CoW \cite{Sahai15} and traditional information broadcasting. For various
schemes, the probability of URLLC is defined as the probability that all the users in the system receive their messages successfully within the delay requirement.
It is shown by simulation that with inter-cell interference, our proposed scheme is able to achieve a probability of URLLC above $99.99\%$ for a much larger rate regime as compared to all the existing URLLC schemes.

\subsection{Organization}
The rest of this paper is organized as follows.
Section \ref{sec:System Model} describes the system model for URLLC.
Section \ref{sec:Proposed Approach for Ultra-Reliable Communications} introduces
the D2D-based two-phase transmission protocol for URLLC.
Section \ref{sec:leader selection based Beamforming Design} describes the
corresponding leader selection based beamforming design.
Section \ref{sec:Benchmark Schemes for Ultra-Reliable Communications} introduces
some benchmark schemes.
Section \ref{sec:Numerical Results} provides the numerical
simulation results pertaining to performance comparison between our proposed scheme and benchmark
schemes. Finally, Section \ref{sec:Conclusion} concludes
this paper.

\section{System Model}\label{sec:System Model}

Consider the downlink communication in one cell (factory) consisting of one BS (controller) and $K$ users (actuators) as shown in Fig. \ref{fig2}.
It is assumed that the BS is equipped with $M$ antennas, and each user is equipped with one single antenna.
It is further assumed that the $K$ users form $N$ disjoint groups based on their geographic locations, while the users
in each group are in close proximity to each other. Let $\mathcal{G}_n$ denote the set of users that belong to group $n$,
and its cardinality $K_n=|\mathcal{G}_n|$ denote the number of users in this group, $n=1,\cdots,N$, respectively. Note that each user
belongs to only one group, thus $\mathcal{G}_n \bigcap \mathcal{G}_j=\emptyset$ if $n\neq j$, and $\sum_{n=1}^NK_n=K$. In practice, the BS can decide how to group the users based on its knowledge of user locations, and send this information to all the users such that each user is aware of which group it belongs to. For convenience, in this paper we assume that user grouping is already done at the network planning stage and also the user grouping information is known to the users.

The downlink channel from the BS to the $k$th user in group $n$ is denoted by $\mv{h}_{k,n}\in \mathbb{C}^{M\times 1}$, $k=1,\cdots,K_n$,
$n=1,\cdots,N$, while the channel from the $i$th user in group $j$ to the $k$th user in group $n$ is denoted by $\tilde{h}_{k,n,i,j}\in \mathbb{C}$,
$\forall (k,n)\neq (i,j)$. This paper adopts a block-fading model, in which all the channels
follow independent quasi-static flat-fading within a block of coherence time, where $\mv{h}_{k,n}$'s and $\tilde{h}_{k,n,i,j}$'s remain constant, but vary independently from block to block.
For convenience, it is assumed that the coherence time and bandwidth of $\mv{h}_{k,n}$'s are the same as those of $\tilde{h}_{k,n,i,j}$'s. They are denoted by $T$ second and $B$ Hz, respectively.
It is further assumed that the downlink channels $\mv{h}_{k,n}$'s are perfectly known at the BS, but the channels between the users $\tilde{h}_{k,n,i,j}$'s are not known at the BS. At last,
we assume that for any user $k$ in group $n$, it knows its downlink channel $\mv{h}_{k,n}$ and the channels from other users to it, i.e., $\tilde{h}_{k,n,i,j}$'s, $\forall (i,j)\neq (k,n)$,
for information decoding.%\footnote{In a TDD system, each user can send one pilot sequence such that the BS and the other users can estimate the channels from this user to them. Then, the BS
%can feedback the downlink channels to the corresponding users.}

For URLLC, let $\tau$ denote the delay requirement for all the users, which in general is much smaller than the channel coherence time, i.e., $\tau<T$.
Furthermore, let $\Omega_{k,n}$ and $D_{k,n}$ denote the set and the number of information bits that need to be conveyed to the $k$th user in group $n$ within the delay requirement, i.e., $\tau$, respectively.
The core question for ultra-reliable communications in this scenario is the following: How to design a protocol such that each of the $K$ users can receive its
messages with a very low decoding error probability within $\tau$ seconds?

\section{Proposed Protocol for Ultra-Reliable Communications}\label{sec:Proposed Approach for Ultra-Reliable Communications}
In this section, we propose a D2D-based two-phase transmission protocol to achieve ultra-reliable communications for the $K$ users located in $N$ groups. We assume that
$\tilde{h}_{k,n,i,j}$ is very strong if $n=j$ since the users in the same group are close to each other, but relatively weak otherwise. As a result, the users in the same group can form a D2D network
in which the communications can be made reliable. The D2D-based two-phase transmission protocol is briefly outlined as follows: in the first phase with a duration
of $\tau_1<\tau$, the BS combines each group's messages together, i.e., $\Omega^{(n)}=\bigcup_{k=1}^{K_n}\Omega_{k,n}$ with $\sum_{k=1}^{K_n}D_{k,n}$ bits information, $\forall n$, and sends $\Omega^{(n)}$'s to the corresponding groups simultaneously via multi-group multicasting; in the second phase with a duration of $\tau_2=\tau-\tau_1$, the users that decode the information successfully in the first phase can help relay the messages to the other users in the same group via the D2D network. Note that under the proposed protocol, each user not only decodes its own messages, but also receives its neighbors' messages, since the successful users in Phase I need to relay other users' messages in the same group in Phase II. In the following, we elaborate this protocol in details.

\subsection{Phase I}
In the first phase with a duration of $\tau_1$ seconds, let $s_n$ denote
the combined symbol intended for all the users in group $n$, which is modeled as a circularly symmetric complex Gaussian (CSCG) random
variable with zero-mean and unit-variance, i.e., $s_n\sim \mathcal{CN}(0,1)$, $\forall n$. Then, the transmit signal of the BS in Phase I is expressed as
\begin{align}\label{eqn:transmit signal Phase I}
\mv{x}^{\rm (I)}=\sum\limits_{n=1}^N\mv{w}_n^{\rm (I)}s_n,
\end{align}where $\mv{w}_n^{\rm (I)}\in \mathbb{C}^{M\times 1}$ denotes the transmit beamformer for the combined user massages of group $n$. Note that messages of different groups are spatially multiplexed. Suppose that
the BS has a transmit power constraint $P_{{\rm BS}}$; from (\ref{eqn:transmit signal Phase I}), we thus have
\begin{align}\label{eqn:power constraint Phase I}
\sum\limits_{n=1}^N \|\mv{w}_n^{({\rm I})}\|^2\leq P_{{\rm BS}}.
\end{align}

The received signal of the $k$th user in group $n$ in Phase I is expressed as
\begin{align}\label{eqn:receive signal in Phase II}
y_{k,n}^{({\rm I})}&=\mv{h}_{k,n}^T\mv{x}^{\rm (I)}+z_{k,n}^{\rm (I)}\nonumber \\ &=\mv{h}_{k,n}^T\mv{w}_n^{\rm (I)}s_n\hspace{-2pt}+\hspace{-2pt}\mv{h}_{k,n}^T\sum\limits_{j\neq n}\mv{w}_j^{\rm (I)}s_j\hspace{-2pt}+\hspace{-2pt}z_{k,n}^{\rm (I)}, ~ \forall k, ~ \forall n,
\end{align}where $z_{k,n}^{\rm (I)}\in \mathcal{CN}(0,I_{k,n}^{({\rm I})})$ denotes the superposition of the additive white Gaussian noise (AWGN) and the inter-cell interference at the $k$th user in group $n$ in Phase I, with a power $I_{k,n}^{({\rm I})}$. The SINR of the $k$th user in group $n$ in Phase I is then expressed as
\begin{align}\label{eqn:SINR in Phase I}
\gamma_{k,n}^{({\rm I})}=\frac{|\mv{h}_{k,n}^T\mv{w}_n^{({\rm I})}|^2}{\sum\limits_{j\neq n}|\mv{h}_{k,n}^T\mv{w}_j^{({\rm I})}|^2+I_{k,n}^{({\rm I})}}.
\end{align}

For the first transmission phase, in total there are $\tau_1 B$ symbols available for the BS to perform information multicasting.
Note that although the seminal work \cite{Polyanskiy10} shows that in an AWGN channel, encoding over finite blocklength can result in a penalty on the channel capacity, it is recently shown in \cite{Polyanskiy14} that in a fading channel, the error event is dominated by the outage due to channel fading, rather than the finite blocklength effect.
As a result, in this paper we ignore the effect of finite blocklength coding, and the minimum SINR target required to convey $\sum_{k=1}^{K_n}D_{k,n}$ bits messages to any user in group $n$ using $\tau_1 B$ symbols is then expressed as
\begin{align}\label{eqn:SINR requirment}
\bar{\gamma}_n^{({\rm I})}=2^{\frac{\sum\limits_{k=1}^{K_n}D_{k,n}}{\tau_1 B}}-1, ~~~ \forall n.
\end{align}

Then, we define an indicator function for each user as follows:
\begin{align}\label{eqn:indicator Phase I}
\phi_{k,n}^{({\rm I})}=\left\{\begin{array}{ll}1, & {\rm if} ~ \gamma_{k,n}^{\rm (I)}\geq \bar{\gamma}_n^{({\rm I})}, \\ 0, & {\rm if}  ~ \gamma_{k,n}^{\rm (I)}< \bar{\gamma}_n^{({\rm I})}. \end{array}\right.
\end{align}

\begin{definition}\label{definition1}
The $k$th user in group $n$ is defined as a \emph{leader} of group $n$ if it can decode the messages $\Omega^{(n)}$ in the first transmission phase, i.e., $\phi_{k,n}^{({\rm I})}=1$.
Moreover, the set of leaders in group $n$ is defined as $\Phi_n^{({\rm I})}=\{k:\phi_{k,n}^{({\rm I})}=1\}$, $n=1,\cdots,N$.
\end{definition}

\subsection{Phase II}\label{sec:Phase II}
In the second phase over the remaining $\tau_2$ seconds, the leaders in each group relay the messages to the unsuccessful
users in their respective group via the local D2D networks. To ensure that all the leaders in one group transmit the same message without generating intra-group interference, we can use one of the following two strategies. In the first strategy, each leader can re-transmit the entire data packet for all the users in its group (including the data for those who already decoded their data). In this case, $\sum_kD_{k,n}$ bits are transmitted by the leaders of group $n$ over $\tau_2$ seconds in Phase II. As an alternative strategy, if each leader is made aware of which other users have successfully decoded their packet (i.e., which other users become leaders), they can subtract the messages of all the leaders from the data packet, re-encode the rest of the messages together, and transmit the newly combined packet in the second phase. Since the leaders' messages $\bigcup_{k\in \Phi_n^{({\rm I})}}\Omega_{k,n}$ are not encoded, only $\sum_{k\notin \Phi_n^{({\rm I})}}D_{k,n}$ information bits need to be transmitted by the leaders of group $n$ over $\tau_2$ seconds in Phase II.

This paper advocates the first transmission strategy mentioned above in Phase II. Although the second strategy transmits fewer number of information bits in Phase II, it requires more overhead for the feedback of control information after Phase I so that each leader is made aware of the other leaders in its group. Specifically, each user needs to feed back one bit message at the end of Phase I to report whether it becomes a leader, then the BS needs to send control messages to the leaders to let them know about the other leaders in their groups. The overhead involved is significant and is likely to overwhelm the benefit of shorter message in Phase II. For this reason, the rest of the paper assumes the use of the first strategy above.

Due to the lack of global CSI of the D2D networks, in this paper we assume that power control
is not performed among users and each leader simply transmits at its full power for relaying the messages. Suppose that all the users possess
a common transmit power constraint $P$. The transmit signal of the $k$th user in group $n$ in the second phase is thus expressed as
\begin{align}\label{eqn:user transmit signal}
x_{k,n}^{({\rm II})}=\phi_{k,n}^{({\rm I})}\sqrt{P}s_n, ~~~ k=1,\cdots,K_n, ~ n=1,\cdots,N.
\end{align}As a result, in Phase II, the received signal for each unsuccessful user in Phase I is expressed as
\begin{align}\label{eqn:receivd signal phase II}
y_{k,n}^{({\rm II})}=&\sum\limits_{(i,j)\neq (k,n)}\tilde{h}_{k,n,i,j}x_{i,j}^{({\rm II})}+z_{k,n}^{\rm (II)} \nonumber
\\ =&\sum\limits_{i\neq k}\tilde{h}_{k,n,i,n}\phi_{i,n}^{({\rm I})}\sqrt{P}s_n +\sum\limits_{j\neq n}\sum\limits_{i=1}^{K_j}\tilde{h}_{k,n,i,j}\phi_{i,j}^{({\rm I})}\sqrt{P}s_j\nonumber \\ & ~~~~~~~~~~~~~~~~~~~~~~~~~~+z_{k,n}^{\rm (II)}, ~ {\rm if} ~ \phi_{k,n}^{({\rm I})}=0,
\end{align}where $z_{k,n}^{\rm (II)}\sim \mathcal{CN}(0,I_{k,n}^{({\rm II})})$ denotes the superposition of the AWGN and the inter-cell interference at the $k$th user in group $n$ in Phase II, with a power $I_{k,n}^{({\rm II})}$.

The corresponding SINR to decode $s_n$ based on $y_{k,n}^{({\rm II})}$ is thus
\begin{align}\label{eqn:SINR phase II}
\gamma_{k,n}^{({\rm II})}=\frac{\left|\sum_{i\neq k}\tilde{h}_{k,n,i,n}\phi_{i,n}^{({\rm I})}\sqrt{P}\right|^2}{\sum_{j\neq n}\left|\sum_{i=1}^{K_j}\tilde{h}_{k,n,i,j}\phi_{i,j}^{({\rm I})}\sqrt{P}\right|^2+I_{k,n}^{({\rm II})}}, ~ {\rm if} ~ \phi_{k,n}^{({\rm I})}=0.
\end{align}

Similar to (\ref{eqn:SINR requirment}), the minimum SINR requirement to relay $\sum_kD_{k,n}$ bits of information using $\tau_2 B$ symbols in Phase II can be expressed as
\begin{align}\label{eqn:SINR requirement Phase II}
\bar{\gamma}_n^{({\rm II})}=2^{\frac{\sum\limits_{k=1}^{K_n}D_{k,n}}{\tau_2 B}}-1, ~~~ \forall n.
\end{align}

Then, we define an indicator function for each user as follows:
\begin{align}\label{eqn:indicator Phase II}
\phi_{k,n}^{({\rm II})}=\left\{\begin{array}{ll}1, & {\rm if} ~ \gamma_{k,n}^{\rm (I)}< \bar{\gamma}_n^{({\rm I})} ~ {\rm and} ~ \gamma_{k,n}^{\rm (II)}\geq \bar{\gamma}_n^{({\rm II})}, \\ 0, & {\rm otherwise}. \end{array}\right.
\end{align}As a result, $\phi_{k,n}^{({\rm II})}=1$ if an unsuccessful user in Phase I decodes the messages in Phase II, and $\phi_{k,n}^{({\rm II})}=0$ otherwise.

%\begin{remark}
%\textcolor{red}{In this paper, we assume that with the aid of control signals from the BS, each leader is aware of all the leaders in its group such that in Phase II, the same message that merely contains information for the users that have not decoded the message in Phase I is transmitted by the leaders of the same group. However, an alternative strategy is to let all the leaders transmit the same combined message that contains all users' message in their group in Phase I. Then, the overall data bits required in Phase II is $\sum_kD_{k,n}$, rather than $\sum_{k\notin \Phi_n^{({\rm I})}}D_{k,n}$. However, control signals from the BS are not required since the leaders do not need to subtract the other leaders' message when transmitting in Phase II.}
%\end{remark}

Define $\Phi_n^{({\rm II})}=\{k:\phi_{k,n}^{({\rm II})}=1\}$ as the set of users that can decode the messages successfully in group $n$, $n=1,\cdots,N$, in Phase II. Then,
the cardinalities of $\Phi_n^{({\rm I})}$ (Definition \ref{definition1}) and $\Phi_n^{({\rm II})}$, i.e., $|\Phi_n^{({\rm I})}|$ and $|\Phi_n^{({\rm II})}|$, indicate the numbers of users in group $n$ who have decoded their messages successfully in Phase I (leaders) and Phase II, respectively. Moreover, $\sum_{n=1}^N|\Phi_n^{({\rm I})}|+\sum_{n=1}^N |\Phi_n^{({\rm II})}|$ denotes the total number of successful users within the cell after $\tau$ second. We define reliable communications in the whole system as follows.

\begin{definition}\label{definition2}
Given a time slot of duration $\tau$ seconds, if some beamforming vectors $\mv{w}_n^{({\rm I})}$'s satisfying the transmit power constraints (\ref{eqn:power constraint Phase I}) can be found at the BS such that all the users can receive their messages under the proposed two-phase transmission protocol, i.e., $\sum_{n=1}^N|\Phi_n^{({\rm I})}|+\sum_{n=1}^N |\Phi_n^{({\rm II})}|=K$, then ultra-reliable communication is achieved over the $\tau$ seconds. Otherwise, we say that an outage has occurred.
\end{definition}

\subsection{Problem Formulation}
To achieve ultra-reliable communication over any particular duration $\tau$, we need to design the beamforming vectors at the BS to maximize the total number of the successful users, i.e.,
\begin{subequations}\label{eqn:problem P}\begin{align}
\mathop{\mathrm{maximize}}_{\{\mv{w}_n^{({\rm I})}\}} & ~ \sum\limits_{n=1}^N|\Phi_n^{({\rm I})}|+\sum\limits_{n=1}^N|\Phi_n^{({\rm II})}| \\
\mathrm {subject \ to}  & ~ \sum\limits_{n=1}^N \|\mv{w}_n^{({\rm I})}\|^2\leq P_{{\rm BS}}. \label{eqn:constraint 1 problem P}
\end{align}\end{subequations}If the optimal value to the above problem is $K$, then ultra-reliable communication is achieved according to Definition \ref{definition2}.

\section{Leader Selection based Beamforming Design}\label{sec:leader selection based Beamforming Design}
Since in practice it is hard to acquire the CSI of the D2D networks, i.e., $\tilde{h}_{k,n,i,j}$'s, at the BS, in this section, we propose a reformulation of problem (\ref{eqn:problem P}) without assuming any knowledge of $\tilde{h}_{k,n,i,j}$'s.

\subsection{Problem Reformulation}\label{sec:Problem Reformulation}
The proposed two-phase transmission protocol in Section \ref{sec:Proposed Approach for Ultra-Reliable Communications} arises from the observation that if a group has at least one leader in the first phase, then with a very high probability, all the other users in the group would be able to decode the messages successfully over the D2D network in the second phase due to their proximity to the leaders. Motivated by this observation, this paper reformulates the problem by setting the constraint of Phase I so as to ensure that each group has at least one leader. This can be done without knowledge of $\tilde{h}_{k,n,i,j}$'s. Moreover, among all beamforming strategies $\mv{w}_n^{({\rm I})}$'s that yield at least one leader for each group, we choose the one that maximizes the total number of leaders in Phase I, so that fewer users need to satisfy their SINR requirements in Phase II. In this way, we formulate the following beamforming design problem, assuming no knowledge of $\tilde{h}_{k,n,i,j}$'s:
\begin{subequations}\label{eqn:problem phase I}\begin{align}
\mathop{\mathrm{maximize}}_{\{\mv{w}_n^{({\rm I})}\}} & ~ \sum\limits_{n=1}^N|\Phi_n^{({\rm I})}| \\
\mathrm {subject \ to}  & ~  |\Phi_n^{({\rm I})}|\geq 1, ~~~ \forall n, \label{eqn:constraint 1 problem phase I}  \\ & ~ \sum\limits_{n=1}^N \|\mv{w}_n^{({\rm I})}\|^2\leq P_{{\rm BS}}. \label{eqn:constraint 2 problem phase I}
\end{align}\end{subequations}

In problem (\ref{eqn:problem phase I}), $\phi_{k,n}^{({\rm I})}$'s as given in (\ref{eqn:indicator Phase I}) are complicated and discrete functions over the beamforming vectors, which make it challenging to apply optimization technique to solve problem (\ref{eqn:problem phase I}). To tackle the issue arising from the discrete $\phi_{k,n}^{({\rm I})}$'s, let us define $\mv{t}_n^{({\rm I})}=[t_{1,n}^{({\rm I})},\cdots,t_{K_n,n}^{({\rm I})}]^T\in \mathbb{C}^{K_n\times 1}$, $\forall n$. It can then be shown that problem (\ref{eqn:problem phase I}) is equivalent to the following problem:
\begin{subequations}\label{eqn:problem P1}\begin{align}
\mathop{\mathrm{minimize}}_{\{\mv{w}_n^{({\rm I})},\mv{t}_n^{({\rm I})}\}} & ~ \sum\limits_{n=1}^N\|\mv{t}_n^{({\rm I})}\|_0 \label{eqn:objective}\\
\mathrm {subject \ to} & ~ \gamma_{k,n}^{({\rm I})}+t_{k,n}^{({\rm I})}\geq \bar{\gamma}_n^{({\rm I})}, ~ \forall k, ~ \forall n, \label{eqn:constraint 1} \\ & ~ \|\mv{t}_n^{({\rm I})}\|_0\leq K_n-1, ~ \forall n, \label{eqn:constraint 2} \\ & ~ t_{k,n}^{({\rm I})}\geq 0, ~ \forall k, ~ \forall n, \label{eqn:constraint 3} \\ & ~ \sum\limits_{n=1}^N \|\mv{w}_n^{({\rm I})}\|^2\leq P_{{\rm BS}}, \label{eqn:constraint 4}
\end{align}\end{subequations}where $\gamma_{k,n}^{({\rm I})}$'s are given in (\ref{eqn:SINR in Phase I}). The equivalence between problem (\ref{eqn:problem phase I}) and problem (\ref{eqn:problem P1}) is due to the fact that the auxiliary variables $t_{k,n}^{({\rm I})}$'s characterize the gap between the SINR targets and achievable SINRs in Phase I, thus the number of zero $t_{k,n}^{({\rm I})}$'s denotes the number of leaders in Phase I, and constraint (\ref{eqn:constraint 2}) guarantees at least one leader in each group. Such an equivalent transformation based on the auxiliary variables $t_{k,n}^{({\rm I})}$'s results in a continuous problem (\ref{eqn:problem P1}). Moreover, since we optimize the number of zero elements in $\mv{t}_n^{({\rm I})}$'s, sparse optimization techniques can now be used to solve the problem.

The difficulty to solve (\ref{eqn:problem P1}) lies in the multiple cardinality constraints (\ref{eqn:constraint 2}). To ensure that there is at least one leader in each group in Phase I without having to deal with the complicated cardinality constraints (\ref{eqn:constraint 2}), in this paper, we propose the following novel \emph{leader selection based beamforming problem}:
\begin{subequations}\label{eqn:problem equivalent Phase I}\begin{align}
\mathop{\mathrm{minimize}}_{\{\mv{w}_n^{({\rm I})},\mv{t}_n^{({\rm I})}\}} & ~ \sum\limits_{n=1}^N\|\mv{t}_n^{({\rm I})}\|_1+\sum\limits_{n=1}^N\beta_n\sqrt[K_n]{\prod\limits_{k=1}^{K_n}t_{k,n}^{({\rm I})}} \label{eqn:objective equivalent Phase I}\\
\mathrm {subject \ to} & ~ (\ref{eqn:constraint 1}), ~ (\ref{eqn:constraint 3}), ~ (\ref{eqn:constraint 4}), \label{eqn:constraint equivalent 1 Phase I}
\end{align}\end{subequations}where $\beta_n>0$ is the corresponding penalty weight for group $n$, $n=1,\cdots,N$.

Note that in the objective function of problem (\ref{eqn:problem equivalent Phase I}), we use the convex functions $\|\mv{t}_n^{({\rm I})}\|_1$'s to approximate non-convex functions $\|\mv{t}_n^{({\rm I})}\|_0$'s in problem (\ref{eqn:problem P1}) based on standard sparse optimization technique. In addition, we define a new penalty\footnote{In this paper, we set the penalty weight as $\beta_n=2^{K_n}$, $\forall n$, i.e., a higher penalty is introduced to the groups with more users since a leader in such groups can help more users in the second phase.} for each group as $\beta_n\sqrt[K_n]{\prod_{k=1}^{K_n}t_{k,n}^{({\rm I})}}$ in the objective function of problem (\ref{eqn:problem P1}). It is easily observed that for each group $n$, its penalty is zero if at least one element of $\mv{t}_n^{({\rm I})}$ is zero. As a result, the new penalty can lead to the desired sparsity pattern, i.e., at least one zero in $\mv{t}_n^{({\rm I})}$, $\forall n$. Moreover, since the geometric mean $f(\mv{x}=[x_1,\cdots,x_N])=\sqrt[N]{\prod_{n=1}^{N}x_n}$ is a concave function over $\mv{x}$ \cite{Boyd04}, the function $\sum_{n=1}^N\beta_n\sqrt[K_n]{\prod_{k=1}^{K_n}t_{k,n}^{({\rm I})}}$ is concave over $t_{k,n}^{({\rm I})}>0$'s. The objective function of problem (\ref{eqn:problem equivalent Phase I}) is thus the difference of convex functions, and standard techniques such as the successive convex approximation method can be used to solve the problem to a local optimum. In the rest of this section, we give details on how to solve the above leader selection based beamforming design problem.

\subsection{Beamforming Design to Problem (\ref{eqn:problem equivalent Phase I})}\label{sec:Beamforming Design in Phase I}

The new penalty in problem (\ref{eqn:problem equivalent Phase I}) enables us to bypass the complicated cardinality constraints of $\mv{t}_n^{({\rm I})}$'s in problem (\ref{eqn:problem P1}). However, problem (\ref{eqn:problem equivalent Phase I}) is a non-convex problem and it is thus difficult to find its globally optimal solution. On one hand, the penalty in the objective function is non-convex. On the other hand, the SINR constraints given in $(\ref{eqn:constraint 1})$ are also non-convex. In the following, we propose an algorithm based on the successive convex approximation technique that can yield a locally optimal solution to problem (\ref{eqn:problem equivalent Phase I}).

First, we provide a convex upper bound to the objective function of problem (\ref{eqn:problem equivalent Phase I}). Since the function $\sum_{n=1}^N\beta_n\sqrt[K_n]{\prod_{k=1}^{K_n}t_{k,n}^{({\rm I})}}$ is concave over $t_{k,n}^{({\rm I})}>0$'s as illustrated in Section \ref{sec:Problem Reformulation}, its first-order approximation at any given point $\hat{\mv{t}}_n^{({\rm I})}=[\hat{t}_{1,n}^{({\rm I})},\cdots,\hat{t}_{K_n,n}^{({\rm I})}]^T$, $\forall n$:
\begin{multline}\label{eqn:approximation to penalty}
f(\{t_{k,n}^{({\rm I})},\hat{t}_{k,n}^{({\rm I})}\}) =\sum\limits_{n=1}^N\beta_n\sqrt[K_n]{\prod\limits_{k=1}^{K_n}\hat{t}_{k,n}^{({\rm I})}} + \\  \sum\limits_{n=1}^N\frac{\beta_n}{K_n}\sqrt[K_n]{\prod\limits_{k=1}^{K_n}\hat{t}_{k,n}^{({\rm I})}}\left[\frac{1}{\hat{t}_{1,n}^{({\rm I})}},\cdots,\frac{1}{\hat{t}_{K_n,n}^{({\rm I})}}\right](\mv{t}_n^{({\rm I})}-\hat{\mv{t}}_n^{({\rm I})}),
\end{multline}serves as its upper bound.

%\begin{IEEEproof}
%Please refer to Appendix \ref{appendix2}.
%\end{IEEEproof}

Next, we deal with the non-convex SINR constraints (\ref{eqn:constraint 1}). Well-known methods to deal with the multicasting SINR constraints include semidefinite relaxation (SDR) \cite{Luo06} and successive convex approximation \cite{Tran14}. Recently, it is shown in \cite{Lu17} that the successive convex approximation based algorithm in general achieves better performance in multicasting than the SDR-based algorithm when the number of antennas at the BS and the number of devices are large. As a result, in this paper, we adopt the successive convex approximation technique to deal with the non-convex SINR constraints (\ref{eqn:constraint 1}). First, the SINR constraints (\ref{eqn:constraint 1}) can be re-formulated as
\begin{align}\label{eqn:new constraint 1}
\frac{|\mv{h}_{k,n}^T\mv{w}_n^{({\rm I})}|^2}{\bar{\gamma}_n^{({\rm I})}}+I_{k,n}^{({\rm I})}t_{k,n}^{({\rm I})}\geq \sum\limits_{j\neq n}|\mv{h}_{k,n}^T\mv{w}_j^{({\rm I})}|^2+I_{k,n}^{({\rm I})}, ~ \forall k, ~ \forall n.
\end{align}The main issue is that $|\mv{h}_{k,n}^T\mv{w}_n^{({\rm I})}|^2$ is a convex function, rather than a concave function, over $\mv{w}_n^{({\rm I})}$. We use the first-order approximation to provide
concave upper bounds to $|\mv{h}_{k,n}^T\mv{w}_n^{({\rm I})}|^2$'s as in \cite{Tran14}. Specifically, define
\begin{align}
& a_{k,n}=\mathfrak{R}(\mv{h}_{k,n}^T\mv{w}_n^{({\rm I})}), ~~~ \forall k, ~ \forall n, \label{1}\\
& b_{k,n}=\mathfrak{I}(\mv{h}_{k,n}^T\mv{w}_n^{({\rm I})}), ~~~ \forall k, ~ \forall n. \label{2}
\end{align}Then we have $|\mv{h}_{k,n}^T\mv{w}_n^{({\rm I})}|^2=a_{k,n}^2+b_{k,n}^2$, $\forall k,n$. Given any $\hat{\mv{w}}_n^{({\rm I})}$'s that satisfy the transmit power constraint (\ref{eqn:power constraint Phase I}), we can define $\hat{a}_{k,n}$'s and $\hat{b}_{k,n}$'s as (\ref{1}) and (\ref{2}). Then, the first-order approximation to
$|\mv{h}_{k,n}^T\mv{w}_n^{({\rm I})}|^2$ at this particular point is given by
\begin{multline}\label{eqn:approximation to SINR Phase I}
g_{k,n}(a_{k,n},b_{k,n},\hat{a}_{k,n},\hat{b}_{k,n})  = \hat{a}_{k,n}^2+\hat{b}_{k,n}^2\\+2[\hat{a}_{k,n},\hat{b}_{k,n}]\left(\left[\begin{array}{c}a_{k,n}\\b_{k,n}\end{array}\right]\hspace{-2pt}-\hspace{-2pt}\left[\begin{array}{c}\hat{a}_{k,n}\\ \hat{b}_{k,n}\end{array}\right]\right), ~ \forall k, n.
\end{multline}To summarize, by introducing the auxiliary variables $a_{k,n}$'s and $b_{k,n}$'s, given any fixed $\hat{\mv{w}}_n^{({\rm I})}$'s, we can use the following convex constraints to approximate the non-convex constraints (\ref{eqn:new constraint 1}):
\begin{multline}\label{eqn:approximated constraint Phase I}
\frac{g_{k,n}(a_{k,n},b_{k,n},\hat{a}_{k,n},\hat{b}_{k,n})}{\bar{\gamma}_n^{({\rm I})}}+I_{k,n}^{({\rm I})}t_{k,n}^{({\rm I})} \\ \geq \sum\limits_{j\neq n}|\mv{h}_{k,n}^T\mv{w}_j^{({\rm I})}|^2+I_{k,n}^{({\rm I})}, ~~~ \forall k, ~ \forall n.
\end{multline}

With the above approximations, given any fixed $\hat{\mv{t}}_n^{({\rm I})}$'s and $\hat{\mv{w}}_n^{({\rm I})}$'s, we can solve the following convex problem:
\begin{subequations}\label{eqn:approximated problem Phase I}\begin{align}
\mathop{\mathrm{minimize}}_{\{\mv{w}_n^{({\rm I})},\mv{t}_n^{({\rm I})},\mv{a}_n,\mv{b}_n\}} & ~ \sum\limits_{n=1}^N\|\mv{t}_n^{({\rm I})}\|_1+f(\{t_{k,n}^{({\rm I})},\hat{t}_{k,n}^{({\rm I})}\}) \label{eqn:objective approximated Phase I}\\
\mathrm {subject \ to} \ \ \ & ~ (\ref{1}), ~ (\ref{2}), ~ (\ref{eqn:approximated constraint Phase I}), ~ (\ref{eqn:constraint 3}), ~ (\ref{eqn:constraint 4}), \label{eqn:constraint approximated 1 Phase I}
\end{align}\end{subequations}where $\mv{a}_n=[a_{1,n},\cdots,a_{K_n,n}]^T$, $\mv{b}_{n}=[b_{1,n},\cdots,b_{K_n,1}]^T$, $\forall n$, and $f(\{t_{k,n}^{({\rm I})},\hat{t}_{k,n}^{({\rm I})}\})$ are as given in (\ref{eqn:approximation to penalty}). Since problem (\ref{eqn:approximated problem Phase I}) is a convex problem, it can be globally solved using existing standard package such as CVX \cite{Boyd11}. The successive convex approximation based algorithm to solve the leader selection based beamforming problem, i.e., (\ref{eqn:problem equivalent Phase I}), proceeds by iteratively updating $\hat{\mv{t}}_n^{({\rm I})}$'s and $\hat{\mv{w}}_n^{({\rm I})}$'s (thus $\hat{\mv{a}}_n$'s and $\hat{\mv{b}}_n$'s) based on the solution to problem (\ref{eqn:approximated problem Phase I}). The proposed algorithm is summarized in Algorithm \ref{table1}. The convergence behavior of Algorithm \ref{table1} is guaranteed in the following proposition.

\begin{algorithm}[t]
{\bf Initialization}: Set the initial values for $\hat{\mv{t}}_n^{({\rm I})}$'s and $\hat{\mv{w}}_n^{({\rm I})}$'s and set $l=1$; \\
{\bf Repeat}:
\begin{enumerate}
\item Find the optimal solution to problem (\ref{eqn:approximated problem Phase I}) using CVX as $\{\mv{w}_n^{({\rm I},l)},\mv{t}_n^{({\rm I},l)},\mv{a}_n^{(l)},\mv{b}_n^{(l)}\}$;
\item Update $\hat{\mv{w}}_n^{({\rm I})}=\mv{w}_n^{({\rm I},l)}$ (thus $\hat{\mv{a}}_n=\mv{a}_n^{(l)}$ and $\hat{\mv{b}}_n=\mv{b}_n^{(l)}$), $\hat{\mv{t}}_n^{({\rm I})}=\mv{t}_n^{({\rm I},l)}$, $\forall n$;
\item $l=l+1$.
\end{enumerate}
{\bf Until} convergence
\caption{Proposed Algorithm for Solving Problem (\ref{eqn:problem equivalent Phase I})}
\label{table1}
\end{algorithm}

\begin{proposition}\label{proposition2}
Monotonic convergence of Algorithm \ref{table1} is guaranteed, i.e.,
\begin{multline}\sum\limits_{n=1}^N\|\mv{t}_n^{({\rm I},l+1)}\|_1+\sum\limits_{n=1}^N\beta_n\sqrt[K_n]{\prod\limits_{k=1}^{K_n}t_{k,n}^{({\rm I},l+1)}}  \\ \leq  \sum\limits_{n=1}^N\|\mv{t}_n^{({\rm I},l)}\|_1+\sum\limits_{n=1}^N\beta_n\sqrt[K_n]{\prod\limits_{k=1}^{K_n}t_{k,n}^{({\rm I},l)}},
\end{multline}where $l$ denotes the index of iteration of Algorithm \ref{table1}. Moreover, the converged solution satisfies all the constraints as well as the Karush-Kuhn-Tucker (KKT) conditions of problem (\ref{eqn:problem equivalent Phase I}).
\end{proposition}

\begin{IEEEproof}
Please refer to Appendix \ref{appendix3}.
\end{IEEEproof}

The leader selection based beamforming design for solving problem (\ref{eqn:problem P}) to maximize the total number of active users under the proposed two-phase transmission scheme in Section \ref{sec:Proposed Approach for Ultra-Reliable Communications} and to achieve ultra-reliable communication as defined in Definition \ref{definition2} is summarized as follows. First, we solve problem (\ref{eqn:problem equivalent Phase I}) via Algorithm \ref{table1} and use $\mv{w}_n^{({\rm I})}$'s to calculate $\phi_{k,n}^{({\rm I})}$'s based on (\ref{eqn:indicator Phase I}). Then, we calculate $\phi_{k,n}^{({\rm II})}$'s based on (\ref{eqn:indicator Phase II}). Reliable communication is achieved if $\sum_{n=1}^N|\Phi_n^{({\rm I})}|+\sum_{n=1}^N |\Phi_n^{({\rm II})}|=K$.

\section{Benchmark Schemes for Ultra-Reliable Communications}\label{sec:Benchmark Schemes for Ultra-Reliable Communications}
In this section, we briefly introduce several other potential approaches for URLLC as benchmark schemes. Numerical simulation comparisons
are provided in Section \ref{sec:Numerical Results} to show the performance gain of our proposed scheme over the benchmark schemes.

\subsection{Benchmark Scheme 1: Our Proposed Scheme but without Leader Selection}
First, to illustrate the importance of leader selection in our proposed two-phase transmission scheme, in the following we consider one benchmark scheme in which
the number of leaders is maximized in Phase I without encouraging at least one leader for each group. In this case, problem (\ref{eqn:problem equivalent Phase I}) reduces to
\begin{subequations}\label{eqn:problem Phase I without leader selection}\begin{align}
\mathop{\mathrm{minimize}}_{\{\mv{w}_n^{({\rm I})},\mv{t}_n^{({\rm I})}\}} & ~ \sum\limits_{n=1}^N\|\mv{t}_n^{({\rm I})}\|_0 \label{eqn:objective Phase I without leade selection}\\
\mathrm {subject \ to} & ~ (\ref{eqn:constraint 1}), ~ (\ref{eqn:constraint 3}), ~ (\ref{eqn:constraint 4}). \label{eqn:constraint 1 Phase I without leader selection}
\end{align}\end{subequations}In other words, the penalty for encouraging at least one leader in each group is removed. Problem (\ref{eqn:problem Phase I without leader selection}) can be solved in a similar way as problem (\ref{eqn:problem equivalent Phase I}).

As a remark, the strategy that maximizes the total number of leaders in Phase I is expected to activate more users in the cell-center groups that are close to the BS, because they have strong direct channels and do not suffer from
inter-cell interference. As a result, it is expected that the optimized beamformers for problem (\ref{eqn:problem Phase I without leader selection}) would result in many leaders in the cell-center groups,
while having no leader in the cell-edge groups. In contrast, our proposed scheme tries to activate at least one leader in each group by solving problem (\ref{eqn:problem equivalent Phase I}). It promotes fairness
among different groups in Phase I and thus makes the best use of the D2D network in Phase II, as later verified by numerical results in Section \ref{sec:Numerical Results}.

\subsection{Benchmark Scheme 2: Occupy CoW Protocol \cite{Sahai15}}
The Occupy CoW protocol is proposed in \cite{Sahai15} for achieving URLLC in a similar setup as in this paper. Specifically, Occupy CoW protocol is also
a two-phase transmission protocol where user messages are transmitted from the BS to the users in the first phase and the successful users help relay the messages to the
other users via the D2D network in the second phase. The main difference of Occupy CoW protocol as compared to our proposed protocol lies in the fact that all the users form one group, rather than $N$ groups based on their
geographic locations. As a result, in the first phase, the BS combines all users' messages together, i.e., $\Omega=\bigcup_{k,n}\Omega_{k,n}$, and multicasts this entire message to all the users,
while in the second phase, the successful users in Phase I can help all the unsuccessful users via the D2D network. Note that \cite{Sahai15} only considers the case that the BS has one single antenna.
To make a fair comparison, in the following we briefly introduce how to design the beamforming under the Occupy CoW protocol if the BS has multiple antennas.

Let $s\sim \mathcal{CN}(0,1)$ denote the entire message intended for all the users. The received signal at each user in the first phase is expressed as
\begin{align}\label{eqn:received signal Phase I Occupy CoW}
y_{k,n}^{({\rm I})}=\mv{h}_{k,n}^T\mv{w}^{({\rm I})}s+z_{k,n}^{({\rm I})}, ~~~ \forall k,n,
\end{align}where $\mv{w}^{({\rm I})}\in \mathbb{C}^{M\times 1}$ denotes the multicast beamforming at the BS in Phase I. The corresponding SINR to decode $s$ is
\begin{align}\label{eqn:SINR Phase I Occupy CoW}
\gamma_{k,n}^{({\rm I})}=\frac{|\mv{h}_{k,n}^T\mv{w}^{({\rm I})}|^2}{I_{k,n}^{{\rm (I)}}}, ~~~ \forall k,n.
\end{align}

Similar to (\ref{eqn:SINR requirment}) in Section \ref{sec:Proposed Approach for Ultra-Reliable Communications}, to decode a message of $\sum_{n=1}^N\sum_{k=1}^{K_n}D_{k,n}$ bits using $B\tau_1$ symbols, the identical minimum SINR requirement for all the users is expressed as:
\begin{align}\label{eqn:SINR requirment Occupy CoW}
\bar{\gamma}^{({\rm I})}=2^{\frac{\sum\limits_{n=1}^N\sum\limits_{k=1}^{K_n}D_{k,n}}{B\tau_1}}-1.
\end{align}The indicator function of each user in Phase I then depends on whether its SINR satisfies this common SINR target or not, i.e.,
\begin{align}\label{eqn:indicator Phase I Occupy CoW}
\phi_{k,n}^{({\rm I})}=\left\{\begin{array}{ll}1, & {\rm if} ~ \gamma_{k,n}^{\rm (I)}\geq \bar{\gamma}^{({\rm I})}, \\ 0, & {\rm if}  ~ \gamma_{k,n}^{\rm (I)}< \bar{\gamma}^{({\rm I})}. \end{array}\right.
\end{align}

The users that decode the messages successfully in the first phase can then relay the messages to the other users in the second phase. Similar to Section \ref{sec:Phase II}, each leader only needs to relay the message $\bigcup_{n}\bigcup_{k\notin \Phi_n^{({\rm I})}}\Omega_{k,n}$ with $\sum_{n=1}^N\sum_{k\notin \Phi_n^{({\rm I})}}D_{k,n}$ bits information to the other users, where $\Phi_n^{({\rm I})}$ denotes the set of leaders located in group $n$. The minimum SINR required to send the above message using $B\tau_2$ symbols is
\begin{align}
\bar{\gamma}^{\rm (II)}=2^{\frac{\sum\limits_{n=1}^N\sum\limits_{k\notin \Phi_n^{({\rm I})}}D_{k,n}}{B\tau_2}}-1.
\end{align}

The key to achieve ultra-reliable communications for the Occupy CoW protocol is the assumption that full diversity gain can be achieved in the second phase even though there is no cooperation between the BS and leaders. Specifically, \cite{Sahai15} assumes that if the $k$th user in group $n$ does not decode the messages in Phase I, based on some space-time coding technique, its SINR achieved from the second phase transmission is
\begin{align}\label{eqn:SINR Phase II Occupy CoW}
\gamma_{k,n}^{({\rm II})}=\max\limits_{(i,j)\neq (k,n)} \gamma_{k,n}^{(i,j)} , ~ {\rm if} ~ \phi_{k,n}^{({\rm I})}=0,
\end{align}where
\begin{align}
\gamma_{k,n}^{(i,j)}=\frac{\phi_{i,j}^{({\rm I})}P|\tilde{h}_{k,n,i,j}|^2}{I_{k,n}^{({\rm II})}}, ~~~ \forall (i,j)\neq (k,n),
\end{align}denotes the individual SINRs of the orthogonal signals from the $i$th user in group $j$. Depending on whether its SINR in Phase II satisfies the SINR target or not, the indicator functions of the unsuccessful users in Phase I are defined as
\begin{align}\label{eqn:indicator Phase II Occupy CoW}
\phi_{k,n}^{({\rm II})}=\left\{\begin{array}{ll}1, & {\rm if} ~ \gamma_{k,n}^{\rm (II)}< \bar{\gamma}^{\rm (II)} ~ {\rm and} ~ \gamma_{k,n}^{\rm (II)}\geq \bar{\gamma}^{\rm (II)}, \\ 0, & {\rm otherwise}. \end{array}\right.
\end{align}

Under the Occupy Cow protocol, the objective of designing the beamforming vectors $\mv{w}^{({\rm I})}$ is to maximize the total number of successful users in Phase I such that more leaders can help relay the messages in Phase II, i.e.,
\begin{subequations}\label{eqn:problem Phase I Occupy CoW}\begin{align}
\mathop{\mathrm{minimize}}_{\{\mv{w}^{({\rm I})},\mv{t}_n^{({\rm I})}\}} & ~ \sum\limits_{n=1}^N\|\mv{t}_n^{({\rm I})}\|_0 \label{eqn:objective Phase I Occupy CoW}\\
\mathrm {subject \ to} & ~ \gamma_{k,n}^{({\rm I})}+t_{k,n}^{({\rm I})}\geq \bar{\gamma}^{({\rm I})}, ~ \forall k,n,  \label{eqn:constraint 1 Phase I Occupy CoW} \\
& ~ t_{k,n}^{({\rm I})}\geq 0, ~ \forall k,n, \label{eqn:constraint 2 Phase I Occupy CoW} \\
& ~ \|\mv{w}^{({\rm I})}\|^2\leq P_{{\rm BS}}. \label{eqn:constraint 3 Phase I Occupy CoW}
\end{align}\end{subequations}Similar to Algorithm \ref{table1}, we can use sparse optimization and successive convex approximation techniques to solve problem (\ref{eqn:problem Phase I Occupy CoW}), and then determine whether URLLC is achieved under the Occupy CoW protocol, i.e., whether all the users receive their messages over $\tau$ seconds.

Note that in \cite{Sahai15}, all the users' channels are assumed to have the same distribution since the locations of the users are not considered in the channel model. This paper shows that if in practice the effect of user locations on user channels is considered, we should group the users in close proximity together since the gain of D2D network in Phase II mainly comes from the adjacent users. Note that in Phase I, user grouping leads to multi-group multicast, instead of single-group multicast under the Occupy CoW protocol in \cite{Sahai15}. If there is only one omnidirectional antenna at the BS, Occupy CoW works well since there is no inter-group interference in Phase I if the BS does not separate users into groups in the spatial domain. If the BS is equipped with multiple antennas, however, our proposed scheme in Section \ref{sec:Proposed Approach for Ultra-Reliable Communications} works better since the BS can adjust $N$ beams in Phase I and utilize the spatial multiplexing gain to activate one leader in each group as long as $M>N$. Note that for Occupy CoW, the BS only designs one beam in Phase I to satisfy the SINR targets of all the users, which is hard. The increased degree of freedom, i.e., optimization of $N$ beams rather than one beam, ensures that our proposed scheme is more powerful than Occupy CoW, as later verified by numerical results in Section \ref{sec:Numerical Results}.

\subsection{Benchmark Scheme 3: Modified Occupy CoW Protocol with Leader Selection}
Leader selection introduced in Section \ref{sec:leader selection based Beamforming Design} can be applied in Occupy CoW as well to improve its performance, if the users are geographically located in separate groups.. In the first phase, although all users' messages are combined together, we can design $\mv{w}^{({\rm I})}$ to activate at least one
leader in each geographical group. The considered problem is thus formulated as
\begin{subequations}\label{eqn:problem Phase I Occupy CoW with leader selection}\begin{align}
\mathop{\mathrm{minimize}}_{\{\mv{w}^{({\rm I})},\mv{t}_n^{({\rm I})}\}} & ~ \sum\limits_{n=1}^N\|\mv{t}_n^{({\rm I})}\|_0+\sum\limits_{n=1}^N\beta_n\sqrt[K_n]{\prod\limits_{k=1}^{K_n}t_{k,n}^{({\rm I})}} \label{eqn:objective Phase I Occupy CoW with leader selection}\\
\mathrm {subject \ to} & ~ (\ref{eqn:constraint 1 Phase I Occupy CoW}), ~ (\ref{eqn:constraint 2 Phase I Occupy CoW}), ~ (\ref{eqn:constraint 3 Phase I Occupy CoW}). \label{eqn:constraint 1 Phase I Occupy CoW with leader selection}
\end{align}\end{subequations}This problem can be solved similarly as problem (\ref{eqn:problem equivalent Phase I}).

For Occupy CoW, combining all users' messages together significantly increases the SINR targets of all the users as compared to our proposed scheme, thus leads to a reduction of the number of leaders in Phase I. In Phase II, an geographically isolated group without a leader nevertheless cannot rely on the far away leaders in other groups. As a result, even if leader selection is considered in Occupy CoW, its probability of reliable communications is much lower than our proposed scheme, as later verified by numerical results in Section \ref{sec:Numerical Results}.

\subsection{Benchmark Scheme 4: One-Phase Transmission Protocol with Broadcasting}
The proposed scheme in Section \ref{sec:Proposed Approach for Ultra-Reliable Communications} and Occupy CoW proposed in \cite{Sahai15} both advocate a two-phase transmission for achieving ultra-reliable communications by utilizing the D2D network in the second phase. In the rest of this section, we introduce possible approaches with one-phase transmission where only the downlink communication from the BS to the users is considered. First, consider the case of broadcasting. In this case, the received signal of the $k$th user in group $n$ is
\begin{multline}\label{eqn:received signal broadcast}
y_{k,n}=\mv{h}_{k,n}^T\mv{w}_{k,n}s_{k,n} \\ +\mv{h}_{k,n}^T\sum\limits_{(i,j)\neq (k,n)}\mv{w}_{i,j}s_{i,j}+z_{k,n}, ~~~ \forall k,n,
\end{multline}where $s_{k,n}\sim \mathcal{CN}(0,1)$ denotes the message for the $k$th user in group $n$, $\mv{w}_{k,n}\in \mathbb{C}^{M \times 1}$ denotes the corresponding beamforming vector, and
$z_{k,n}\sim \mathcal{CN}(0,I_{k,n})$ denotes the superposition of the AWGN and inter-group interference, with a power $I_{k,n}$. The SINR for decoding $s_{k,n}$ is thus
\begin{align}\label{eqn:SINR broadcast}
\gamma_{k,n}=\frac{|\mv{h}_{k,n}^T\mv{w}_{k,n}|^2}{\sum\limits_{(i,j)\neq (k,n)}|\mv{h}_{k,n}^T\mv{w}_{i,j}|^2+I_{k,n}}, ~~~ \forall k,n.
\end{align}

Note that the minimum SINR target to deliver $D_{k,n}$ bits information using $B\tau$ symbols is:
\begin{align}\label{eqn:SINR requirment broadcast}
\tilde{\gamma}_{k,n}=2^{\frac{D_{k,n}}{B\tau}}-1, ~~~ \forall k,n.
\end{align}Then, we can design the beamforming vectors $\mv{w}_{k,n}$'s to maximize the number of users whose SINRs satisfy their SINR targets, i.e.,
\begin{subequations}\label{eqn:problem broadcast}\begin{align}
\mathop{\mathrm{minimize}}_{\{\mv{w}_{k,n},t_{k,n}\}} & ~ \sum\limits_{n=1}^N\sum\limits_{k=1}^{K_n}|t_{k,n}|_0 \label{eqn:objective broadcast}\\
\mathrm {subject \ to} & ~ \gamma_{k,n}+t_{k,n}\geq \tilde{\gamma}_{k,n}, ~ \forall k,n,  \label{eqn:constraint 1 broadcast} \\
& ~ t_{k,n}^{({\rm I})}\geq 0, ~ \forall k,n, \label{eqn:constraint 2 broadcast} \\
& ~ \sum\limits_{n=1}^N\sum\limits_{k=1}^{K_n}\|\mv{w}_{k,n}\|^2\leq P_{{\rm BS}}. \label{eqn:constraint 3 broadcast}
\end{align}\end{subequations}

Similar to \cite{Shamai06}, we can transform the SINR constraint (\ref{eqn:constraint 1 broadcast}) into the following convex form:
\begin{multline}\label{eqn:SINR SOCP}
\frac{\mv{h}_{k,n}^T\mv{w}_{k,n}}{\tilde{\gamma}_{k,n}}+I_{k,n}t_{k,n}  \\ \geq  \sqrt{\sum\limits_{(i,j)\neq (k,n)}|\mv{h}_{k,n}^T\mv{w}_{i,j}|^2+I_{k,n}}, ~~~ \forall k,n.
\end{multline}Then, via relaxing $|t_{k,n}|_0$'s by $|t_{k,n}|_1$'s in the objective function (\ref{eqn:objective broadcast}), we can design the beamforming vectors by solving the
following convex problem:
\begin{subequations}\label{eqn:problem broadcast relax}\begin{align}
\mathop{\mathrm{minimize}}_{\{\mv{w}_{k,n},t_{k,n}\}} & ~ \sum\limits_{n=1}^N\sum\limits_{k=1}^{K_n}|t_{k,n}|_1 \label{eqn:objective broadcast relax}\\
\mathrm {subject \ to} & ~ (\ref{eqn:SINR SOCP}), ~ (\ref{eqn:constraint 2 broadcast}), ~ (\ref{eqn:constraint 3 broadcast}). \label{eqn:constraint 1 broadcast relax}
\end{align}\end{subequations}URLLC via broadcasting is achieved if with the obtained beamforming vectors, all the users satisfy their SINR targets.

As a remark, in most use cases for URLLC the rate requirement of each user is very low, thus even if we combine each group's messages together, the SINR requirement for each group is still reasonable. In this case, intuitively, with $M$ antennas at the BS, approximately $M$ leaders and thus $M$ groups with one leader in each group can be supported under our proposed scheme. On the other hand, information broadcasting can only support approximately $M$ users in toal, i.e., URLLC is not possible if $K\gg M$. With a reliable D2D network, our proposed scheme can achieve URLLC even when $K\gg M$ as long as the number of groups satisfy $N<M$.

\subsection{Benchmark Scheme 5: One-Phase Transmission Protocol with TDMA}
Another strategy to convey $\Omega_{k,n}$ to the $k$th user in group $n$, $\forall k,n$, is TDMA. Specifically, when the $k$th user in group $n$ is scheduled, the BS can implement a maximal-ratio transmission (MRT) beamforming, and the received signal at the user is
\begin{align}\label{eqn:received signal TDMA}
y_{k,n}=\mv{h}_{k,n}^T\frac{\sqrt{p_{k,n}}\mv{h}_{k,n}^\ast}{\|\mv{h}_{k,n}\|}s_{k,n}+z_{k,n}, ~~~ \forall k,n,
\end{align}where $p_{k,n}$ denotes the power to transmit the message $\Omega_{k,n}$. The corresponding SINR at the user is
\begin{align}\label{eqn:SINR TDMA}
\gamma_{k,n}=\frac{p_{k,n}\|\mv{h}_{k,n}\|^2}{I_{k,n}}, ~~~ \forall k,n.
\end{align}

Note that with TDMA, $D_{k,n}$ bits of information needs to be transmitted to the $k$th user in group $n$ with only $B\tau/K$ symbols. The corresponding minimum SINR target for each user is:
\begin{align}\label{eqn:SINR requirment TDMA}
\hat{\gamma}_{k,n}=2^{\frac{KD_{k,n}}{B\tau}}-1, ~~~ \forall k,n.
\end{align}Then, the transmit power to satisfy each user's SINR target is
\begin{align}
p_{k,n}^\ast=\frac{\hat{\gamma}_{k,n}I_{k,n}}{\|\mv{h}_{k,n}\|^2}, ~~~ \forall k,n.
\end{align}If the total transmit power at the BS satisfies the transmit power constraint, i.e.,
\begin{align}
\sum_{n=1}^N\sum_{k=1}^{K_n}p_{k,n}^\ast\leq P_{{\rm BS}},
\end{align}then all the users can be supported via TDMA, and thus URLLC is achieved.

In a massive connectivity scenario, the number of users, i.e., $K$, can be large. If each user is only allocated $\tau/K$ seconds for information transmission, the resulting SINR target as shown in (\ref{eqn:SINR requirment TDMA}) is quite high. As a result, as later shown in Section \ref{sec:Numerical Results}, TDMA cannot achieve reliable communications in general.

\subsection{Benchmark Scheme 6: One-Phase Transmission Protocol with Multi-Group Multicasting}
Moreover, similar to the first phase in our proposed scheme, we can also apply multi-group multicasting to convey the message $\Omega^{(n)}$ with $\sum_{k=1}^{K_n}D_{k,n}$ bits to all the users in group $n$, $\forall n$, but using all the $B\tau$ symbols in one shot. The SINR of each user is given in (\ref{eqn:SINR in Phase I}). However, since the transmission time is doubled compared to our proposed two-phase transmission scheme, the SINR target for each group is reduced. Specifically, the minimum SINR requirement for the users in group $n$ is:
\begin{align}\label{eqn:SINR requirment multicast one phase}
\tilde{\gamma}_n=2^{\frac{\sum\limits_{k=1}^{K_n}D_{k,n}}{B\tau}}-1, ~~~ \forall n.
\end{align}We can proceed to solve problem (\ref{eqn:problem Phase I without leader selection}) in Benchmark Scheme 1 with the new SINR targets $\tilde{\gamma}_n$'s. If one beamforming solution can be found
such that all the users can decode their messages, then URLLC is achieved for this scheme.

However, similar to Benchmarks 4 and 5, the lack of utilization of the reliable D2D network makes it difficult for this one-phase scheme to guarantee the performance of cell-edge users who experience strong inter-cell interference.

\section{Numerical Results}\label{sec:Numerical Results}
This section presents the numerical results to verify the effectiveness of the proposed two-phase transmission protocol in Sections \ref{sec:Proposed Approach for Ultra-Reliable Communications}
and \ref{sec:leader selection based Beamforming Design} for URLLC as compared to the benchmark schemes introduced in Section \ref{sec:Benchmark Schemes for Ultra-Reliable Communications}.

\begin{figure}[t]
\begin{center}
\scalebox{0.4}{\includegraphics*{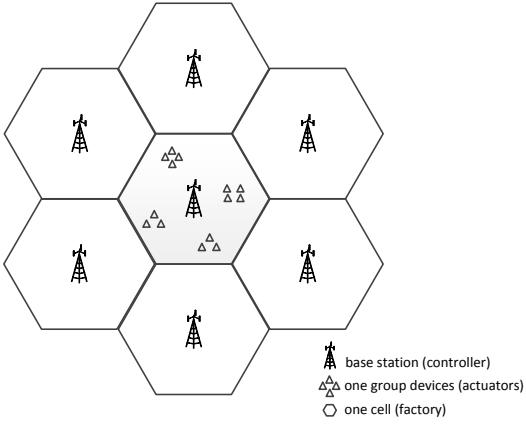}}
\end{center}
\caption{The cellular-based model for industrial automation: the cell, BS, and user act as factory, controller, and actuator, respectively, in industrial automation. In this setup, the center cell is the reference cell of interests, while the other 6 cells generate inter-cell interference to the users in the reference cell.}\label{fig1}
\end{figure}

The simulation setup is as follows. As shown in Fig. \ref{fig1}, we consider a network consisting of 7 cells in a wrapped around topology, in which the center cell is the reference cell---the performance of which we are interested in, and the other 6 adjacent cells generate inter-cell interference to the users in the reference cell.\footnote{In our simulation, for each channel realization, we first randomly generate the other 6 cells' beamformers at the BSs in Phase I and leader locations in Phase II. Then, we calculate the inter-cell interference for the reference cell and design its beamforming accordingly as shown in Section \ref{sec:leader selection based Beamforming Design}. In other words, only the reliability of the reference cell is considered.} This cellular-based topology is a proper model for industrial automation: each cell, BS, and device can be viewed as the factory, controller, and actuator, respectively, in which each factory consists of several groups of actuators. As a result, the simulation results shown in this section are valid in an industrial automation topology.

In the simulation, the cell radius is $500$m. It is assumed that there are $N=6$ groups in the reference cell, while each group consists of $8$ users, i.e, $K=48$. The center of each group is randomly located in a doughnut shape of inner radius $R_{{\rm inner}}$ and outer radius $R_{{\rm outer}}$. Moreover, each group covers an area with a radius of $20$m, and its users are randomly located in the area covered by this group. The BS is assumed to be equipped with $M=8$ antennas. The downlink channel from the BS to the $k$th user in group $n$ is modeled as $\mv{h}_{k,n}= \eta_{k,n}\mv{g}_{k,n}$, where $\eta_{k,n}$ denotes the path-loss component, and $\mv{g}_{k,n}\sim \mathcal{CN}(\mv{0},\mv{I})$ denotes the Rayleigh fading component. Moreover, the path-loss component is modeled as $-128.1-36.7\log_{10}(d_{k,n})$ in dB, where $d_{k,n}$ in km denotes the distance from the $k$th user in group $n$ to the BS. For the communications between the users over the D2D network,
if two users are in the same group, we model their channel as Rician fading; otherwise, we model their channel as Rayleigh fading. Specifically, the channel from the $i$th user in group $n$ to the $k$th user in the same group
is modeled as $\tilde{h}_{k,n,i,n}=\eta_{k,n,i,n}(\sqrt{\delta/(\delta+1)}\hat{g}_{k,n,i,n}+ \sqrt{1/(\delta+1)}\tilde{g}_{k,n,i,n})$, where $\eta_{k,n,i,n}$ denotes the path-loss component, $\hat{g}_{k,n,i,n}$ with $|\hat{g}_{k,n,i,n}|^2=1$ denotes the line-of-sight (LOS) deterministic component, $\tilde{g}_{k,n,i,n}\sim \mathcal{CN}(0,1)$ denotes the Rayleigh fading component, and $\delta$ denotes the Rician factor specifying the power ratio between the LOS and fading component. In this paper, we set $\delta=4$ and characterize the path-loss component as $-76.8-18.7\log_{10}(d_{k,n,i,n})$ in dB according to the indoor channel model in \cite{Channel}, where $d_{k,n,i,n}$ denotes the distance between the two users. Moreover, the channel from the $i$th user in group $j$ to the $k$th user in another group $n$ is modeled as $\tilde{h}_{k,n,i,j}=\eta_{k,n,i,j}\tilde{g}_{k,n,i,j}$, where the path-loss component is modeled as $-128.1-36.7\log_{10}(d_{k,n,i,j})$ in dB, and $\tilde{g}_{k,n,i,j}\sim \mathcal{CN}(0,1)$ denotes the Rayleigh fading component. The total bandwidth used is assumed to be $B=100$kHz for both the downlink channels $\mv{h}_{k,n}$'s and D2D channels $\tilde{h}_{k,n,i,j}$'s. We assume flat fading across this bandwidth $B$. The delay requirement for the communication is $\tau=1$ms. As a result, there are in total $B\tau=100$ transmit symbols available for delivering the messages $\Omega_{k,n}$'s to each user. Moreover, we set $\tau_1=0.75$ms in Phase I and $\tau_2=0.25$ms in Phase II. The motivation for allocating more time to Phase I is to reduce its SINR requirement (see (\ref{eqn:SINR requirment})) such that there is at least one leader in each group. The transmit power constraint at the BS is $P_{{\rm BS}}=43$dBm, and at the user is $P=23$dBm. The power spectral density of the AWGN at the users is assumed to be $-169$dBm/Hz. For convenience, it is assumed that the message sizes of all the users are the same, i.e., $D_{k,n}=D$, $\forall k,n$. In the simulation, we generate $10,000$ channel realizations and for each realization, we measure whether all the users decode their messages within $\tau=1$ms for each investigated scheme. The probability of URLLC for each scheme is then defined as the percentage of instances achieving reliable communication.

\subsection{Performance Comparison between Proposed Scheme and Benchmark Schemes when $D=22$}
In this numerical example, we assume that each user requires a $22$-bit message, i.e., $D=22$. Moreover, we assume that $R_{{\rm inner}}=250$m and $R_{{\rm outer}}=350$m, i.e., the center of each group is randomly located in a doughnut shape of inner radius $250$m and outer radius $350$m. The probabilities of reliable communications and the averaged numbers of users that decode their messages successfully under the proposed scheme versus Benchmark Schemes 1 -- 6 are given in Tables \ref{table4} and \ref{table5}, respectively. It is observed that our proposed scheme can achieve a probability of reliable communications above $99.99\%$ (no outage is observed in $10,000$ channel realizations) in this setup, which is much higher than the benchmark schemes. Specifically, if leader selection is not considered in our proposed scheme, i.e., Benchmark Scheme 1, the probability of reliable communications is 0, although the averaged number of successful users is $32.241$. It is worth noting that under the proposed scheme with leader selection, all the $8$ groups have at least one leader after Phase I in all the $10,000$ channel realizations, while under Benchmark Scheme 1 without leader selection, on average only $4.2$ groups have at least one leader. This verifies the effectiveness of the leader selection based beamforming design in our proposed scheme. Moreover, for the Occupy Cow protocol, no user can decode the message for both the cases without and with leader selection, i.e., Benchmark Schemes 2 and 3, since the SINR requirement for multicasting $KD=1056$ bits to the users using $\tau_1B=75$ symbols is too high (about $42$dB). Finally, for the one-phase transmission scheme, information broadcasting, i.e., Benchmark Scheme 4, can achieve reliable communications with a probability of $11.60\%$, while the probabilities of reliable communications achieved by TDMA and information multicasting, i.e., Benchmark Schemes 5 and 6, are both zero.

\begin{table}
\centering
\begin{center}
\caption{Probability of reliable communications when $D=22$ bits} \label{table4}
{\small
\begin{tabular}{|c|c|}
\hline Proposed Scheme & $> 99.99\%$ \\
\hline
Benchmark Scheme 1 & $0$ \\
\hline
Benchmark Scheme 2 & $0$ \\
\hline
Benchmark Scheme 3 & $0$ \\
\hline
Benchmark Scheme 4 & $11.60\%$ \\
\hline
Benchmark Scheme 5 & $0$ \\
\hline
Benchmark Scheme 6 & $0$ \\
\hline
\end{tabular}
}
\end{center}
\end{table}

\begin{table}
\centering
\begin{center}
\caption{Averaged total number of users that decode their messages successfully when $D=22$ bits} \label{table5}
{\small
\begin{tabular}{|c|c|}
\hline Proposed Scheme & $48.0$ \\
\hline
Benchmark Scheme 1 & $32.2$ \\
\hline
Benchmark Scheme 2 & $0$ \\
\hline
Benchmark Scheme 3 & $0$ \\
\hline
Benchmark Scheme 4 & $45.1$ \\
\hline
Benchmark Scheme 5 & $0.014$ \\
\hline
Benchmark Scheme 6 & $11.5$ \\
\hline
\end{tabular}
}
\end{center}
\end{table}

\subsection{How Many Bits can Be Transmitted Reliably?}
In this example, we study the effect of transmission rate, which is determined by the size of message per user, i.e., $D$, the latency requirement $1$ms and the bandwidth used, i.e., $100$kHz, on the probability of reliable communications. This result is useful in practice because it indicates how much information can be reliably sent to all the users under each scheme. The simulation setup is the same as that of Tables \ref{table4} and \ref{table5}.

\begin{figure}[t]
\begin{center}
\scalebox{0.6}{\includegraphics*{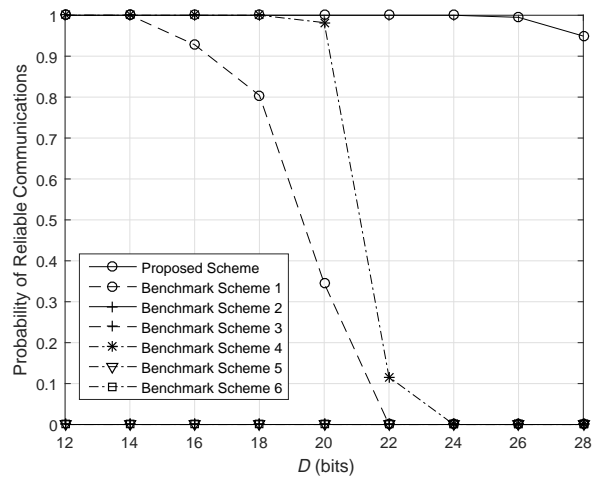}}
\end{center}
\caption{Probabilities of reliable communications versus values of $D$ under different schemes. The center of each group is randomly located in a doughnut shape of inner radius $R_{{\rm inner}}=250$m and outer radius $R_{{\rm outer}}=350$m.}\label{fig3}
\end{figure}

Fig. \ref{fig3} shows the probabilities of reliable communications achieved by different schemes versus various values of $D$. Under our proposed scheme, when $D\leq 24$ bits, outage is not observed in the $10,000$ channel realizations, while when $D=26$ or $28$ bits, $53$ and $523$ outages are observed over $10,000$ channel realizations, respectively. As a result, if the reliability requirement of the communications is $99.99\%$, i.e., no more than $1$ outages in $10,000$ channel realizations, then we can transmit at most $D=24$ bits to each user in this setup.

We remark that in Section \ref{sec:Phase II} we provide an alternative transmission strategy for Phase II in which each leader is made aware of all the other leaders in its group so it can  subtract all the leaders' messages and transmit a shorter packet. We ignore the overhead for the control signals and  measure the reliability achieved by this scheme. It is observed that still at most $D=24$ bits can be transmitted to each user if the reliability requirement of the communications is $99.99\%$. The reason is as follows. As observed in our simulation, when the number of information bits required by the users is $D=24$ bits, there are only about $1-2$ leaders among $8$ users in each group after multi-group multicast beamforming in Phase I. In this case, the message size is not significantly reduced in Phase II even if the leaders are allowed to subtract the messages of all the leaders.

Moreover, it is observed that when $D\leq 18$ bits, reliable communications can be achieved by information broadcasting as well. However, Occupy CoW Protocol cannot achieve reliable communications even when $D= 12$ bits. This is because when $D=12$ bits, the minimum SINR requirement for multicasting $KD=576$ bits to the users using $\tau_1B=75$ symbols is $23.10$dB, which is too high for the cell-edge users. In \cite{Sahai15}, it is shown that if the BS has one antenna, i.e., $M=1$, many users' channels from the BS can suffer from deep fading at any time, and the two-phase Occupy Cow protocol is able to combat the fading and enhance the system reliability by increasing the diversity order in the second phase, if the rate requirement of each user is sufficiently small so that a large number of users can decode the message in the first phase. However, if the BS has multiple antennas, the effect of channel fading on the reliability of the information broadcasting is reduced thanks to the channel diversity. In this case, information broadcasting is more effective than Occupy CoW since combining all user messages together results in an infeasible rate for all the users under Occupy CoW, while broadcasting a very short message to each user would be feasible in many cases via beamforming.

\subsection{Effect of User Topology on System Reliability}
In this subsection, we study the effect of user topology on the reliability of our considered communication system. The simulation setup is the same as that of Fig. \ref{fig3}, except that the center of each group is randomly located in a doughnut shape of inner radius $R_{{\rm inner}}=350$m and outer radius $R_{{\rm outer}}=450$m, i.e., the users suffer from stronger inter-cell interference.

\begin{figure}[t]
\begin{center}
\scalebox{0.6}{\includegraphics*{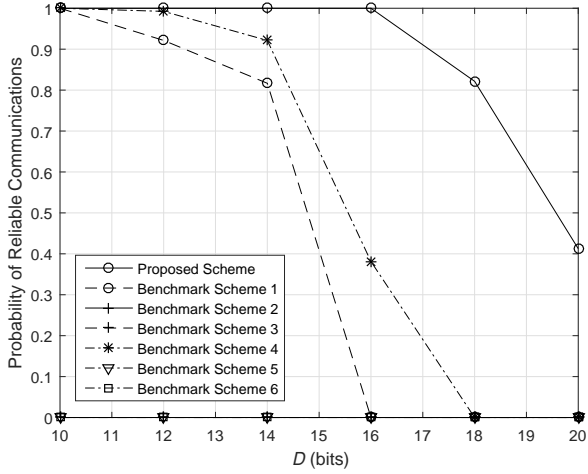}}
\end{center}
\caption{Probabilities of reliable communications versus values of $D$ under different schemes. The center of each group is randomly located in a doughnut shape of inner radius $R_{{\rm inner}}=350$m and outer radius $R_{{\rm outer}}=450$m.}\label{fig4}
\end{figure}

Fig. \ref{fig4} shows the probabilities of reliable communications achieved by different schemes versus values of $D$. By comparing Fig. \ref{fig4} with Fig. \ref{fig3}, it is observed that since in this numerical example the users are located at the cell edge with much stronger inter-cell interference, the system reliability is significantly reduced. Specifically, if the reliability requirement of the communications is $99.99\%$, i.e., no more than $1$ outages in $10,000$ channel realizations, then we can transmit at most $D=16$ bits to each user under our proposed scheme, which is much lower than $D=24$ bits in Fig. \ref{fig3}. This example shows that the system reliability heavily depends on the user distribution. Nevertheless, our scheme still outperforms the other benchmark schemes significantly in this numerical example.

\section{Conclusion and Future Work}\label{sec:Conclusion}
This paper proposes a novel two-phase transmission protocol to fully exploit the D2D transmission for URLLC. Under the proposed protocol, each group's messages are combined
together and multicast to the leaders from the BS in the first phase, while the leaders relay the messages to the other users in their groups in the second phase. Since the D2D networks
are reliable due to the strong channels between the users in the same group, the challenge of our protocol is to select at least one leader for each group in the first phase via a proper
beamforming design at the BS. Utilizing the sparse optimization technique, this paper proposes an efficient algorithm that jointly optimizes the beamforming at the BS and leader selection
in each group. Simulation results show that the proposed algorithm results in a fair leader assignment among groups in the first phase, thus leading to reliable communications
in all the groups in the second phase. Performance comparison to other existing schemes for URLLC is provided to show the effectiveness of the proposed protocol with the leader selection
based beamforming design.

There are a number of directions along which the results of this paper can be further extended. For example, this paper assumes that the user grouping is already done at the network planning stage and known to the BS and users. Future work may study how to group the users based on their locations such that the system reliability can be maximized under our proposed downlink protocol. Moreover, this paper focuses on URLLC in the downlink. Future work may investigate efficient protocols for achieving URLLC in the uplink.

\begin{appendix}
\subsection{Proof of Proposition \ref{proposition2}}\label{appendix3}
First, it can be shown that in the $l$th iteration of Algorithm \ref{table1}, the solution obtained in the $(l-1)$th iteration is also feasible to problem (\ref{eqn:approximated problem Phase I}) given $\hat{\mv{w}}_n^{({\rm I})}=\mv{w}_n^{({\rm I},l-1)}$ (thus $\hat{\mv{a}}_n=\mv{a}_n^{(l-1)}$ and $\hat{\mv{b}}_n=\mv{b}_n^{(l-1)}$) and $\hat{\mv{t}}_n^{({\rm I})}=\mv{t}_n^{({\rm I},l-1)}$, $\forall n$. In other words, $\sum_{n=1}^N\|\mv{t}_n^{({\rm I},l-1)}\|_1+f(\{t_{k,n}^{({\rm I},l-1)},\hat{t}_{k,n}^{({\rm I},l-1)}\})$ is achievable to problem (\ref{eqn:approximated problem Phase I}) in the $l$th iteration. As a result, the optimal objective value of problem (\ref{eqn:approximated problem Phase I}) in the $l$th iteration is no larger than that achieved by the solution $\mv{w}_n^{({\rm I},l-1)}$'s and $\mv{t}_n^{({\rm I},l-1)}$'s, i.e., \begin{multline}
\sum_{n=1}^N\|\mv{t}_n^{({\rm I},l)}\|_1+f(\{t_{k,n}^{({\rm I},l)},\hat{t}_{k,n}^{({\rm I},l-1)}\})\leq \\ \sum_{n=1}^N\|\mv{t}_n^{({\rm I},l-1)}\|_1+f(\{t_{k,n}^{({\rm I},l-1)},\hat{t}_{k,n}^{({\rm I},l-1)}\}).\end{multline}It is worth noting that $f(\{t_{k,n}^{({\rm I},l-1)},\hat{t}_{k,n}^{({\rm I},l-1)}\})=\sum_{n=1}^N\beta_n\sqrt[K_n]{\prod_{k=1}^{K_n}t_{k,n}^{({\rm I},l-1)}}$ according to (\ref{eqn:approximation to penalty}). It is also worth noting that since $\sum_{n=1}^N\beta_n\sqrt[K_n]{\prod_{k=1}^{K_n}t_{k,n}^{({\rm I})}}$ is a concave function, we have
\begin{align}
f(\{t_{k,n}^{({\rm I},l)},\hat{t}_{k,n}^{({\rm I},l-1)}\})\geq \sum_{n=1}^N\beta_n\sqrt[K_n]{\prod_{k=1}^{K_n}t_{k,n}^{({\rm I},l)}}.
\end{align} Then, it follows that
\begin{multline}
\sum\limits_{n=1}^N\|\mv{t}_n^{({\rm I},l)}\|_1+\sum\limits_{n=1}^N\beta_n\sqrt[K_n]{\prod\limits_{k=1}^{K_n}t_{k,n}^{({\rm I},l)}}  \\ \leq  \sum\limits_{n=1}^N\|\mv{t}_n^{({\rm I},l)}\|_1+f(\{t_{k,n}^{({\rm I},l)},\hat{t}_{k,n}^{({\rm I},l-1)}\}) ~~~~~~ \\ \leq  \sum\limits_{n=1}^N\|\mv{t}_n^{({\rm I},l-1)}\|_1+f(\{t_{k,n}^{({\rm I},l-1)},\hat{t}_{k,n}^{({\rm I},l-1)}\})
\\ =  \sum\limits_{n=1}^N\|\mv{t}_n^{({\rm I},l-1)}\|_1+\sum\limits_{n=1}^N\beta_n\sqrt[K_n]{\prod\limits_{k=1}^{K_n}t_{k,n}^{({\rm I},l-1)}}.
\end{multline}Monotonic convergence of Algorithm \ref{table1} is thus proved.

Next, since in Algorithm \ref{table1} we use lower-bound to approximate the non-concave functions in problem (\ref{eqn:problem equivalent Phase I}), as shown in (\ref{1}), (\ref{2}), and (\ref{eqn:approximated constraint Phase I}), any feasible solution to problem (\ref{eqn:approximated problem Phase I}) satisfies all the constraints of problem (\ref{eqn:problem equivalent Phase I}).

Lastly, according to \cite[Theorem 1]{Marks78}, the solution obtained by the successive convex approximation based Algorithm \ref{table1} must satisfy the KKT conditions of problem (\ref{eqn:problem equivalent Phase I}).

\end{appendix}

\end{document}